*Biophysics at the coffee shop: lessons learned working with George Oster*

Oleg Igoshin[1, *], Jing Chen[2, *], Jianhua Xing[3, *], Jian Liu[4, *], Timothy C. Elston[5], Michael Grabe[6], Kenneth S. Kim[7], Jasmine Nirody[8], Padmini Rangamani[9], Sean Sun[10], Hongyun Wang[11], Charles Wolgemuth[12]

1. Departments of Bioengineering, of Biosciences and of Chemistry and Center for Theoretical Biological Physics, Rice University, Houston, TX 77005, USA;
2. Department of Biological Sciences, Virginia Tech, Blacksburg, VA 24061, USA;
3. Department of Computational and Systems Biology, and UPMC-Hillman Cancer Center, School of Medicine, University of Pittsburgh, Pittsburgh, PA, 15261, USA;
4. Department of Cell Biology and Center for Cell Dynamics, School of Medicine, Johns Hopkins University, Baltimore, MD 21205, USA;
5. Department of Pharmacology, School of Medicine, University of North Carolina at Chapel Hill, Chapel Hill, NC 27599, USA;
6. Cardiovascular Research Institute, School of Pharmacy, University of California at San Francisco, San Francisco, CA 94158, USA;
7. Lawrence Livermore National Laboratory, Livermore, CA 94550, USA;
8. Center for Studies in Physics and Biology, Rockefeller University, New York, NY 10065, USA;
9. Department of Mechanical and Aerospace engineering, University of California at San Diego, La Jolla, CA 92093, USA;
10. Department of Mechanical Engineering, Whiting School of Engineering, Johns Hopkins University, Baltimore, MD 21218, USA;
11. Department of Applied Mathematics and Statistics, University of California Santa Cruz, CA 95064
12. Department of Physics, Department of Molecular and Cellular Biology, University of Arizona, Tucson, AZ 85721, USA.

(*) Corresponding: Oleg Igoshin (igoshin@rice.edu); Jing Chen (chenjing@vt.edu); Jianhua Xing (xing1@pitt.edu); Jian Liu (jliu187@jhmi.edu).

**Abstract**

Over the past 50 years, the use of mathematical models, derived from physical reasoning, to describe molecular and cellular systems has evolved from an art of the few to a cornerstone of biological inquiry. George Oster stood out as a pioneer of this paradigm shift from descriptive to quantitative biology not only through his numerous research accomplishments, but also through the many students and postdocs he mentored over his long career. Those of us fortunate enough to have worked with George agree that his sharp intellect, physical intuition and passion for scientific inquiry not only inspired us as scientists but also greatly influenced the way we conduct research. We would like to share a few important lessons we learned from George in honor of his memory and with the hope that they may inspire future generations of scientists.

**Introduction**

The use of mathematical models derived from physical and engineering principles to explain phenomena occurring at the cellular and molecular level goes back well over half a century. Despite some notable successes (such as the Hodgkin–Huxley model of the neuronal action potential (Hodgkin and Huxley, 1952) that earned the 1963 Nobel Prize in Physiology), over those first few decades mathematical models were generally-viewed with skepticism by molecular and cell biologists. However, with the publication of more and more theory-driven papers providing deep biological insights and with the emergence of new experimental techniques that could perturb and measure cellular systems with high spatiotemporal resolution, the tides are slowly turning. It is becoming increasingly common for mathematical modeling to be incorporated into biological and biomedical research, and departments in the biological sciences are increasingly investing in faculty with training in the physical and mathematical sciences.

George Oster's numerous contributions to biophysical theory and modeling were instrumental in making such a paradigm shift possible. While over his 50-year career George's research addressed different biological scales ranging from the macromolecular and cellular levels all the way up to ecology and evolution, every project reflected his distinct approach to modeling. George adopted a reverse-engineering philosophy for each system of interest and strove to find out "how things work" in each case. He was not a big fan of models that are overly simplified for the sake of mathematical tractability, but bore little relation to the biological system under consideration. Instead, he always aimed to solve puzzles raised by experimental observations and to find solutions that were constrained by data and underlying physical principles. To this end, he would find the method most suitable for answering the question at hand. As George liked to put it: *"You want to find the tool for the job, instead of holding a hammer and looking for a nail."*

Those of us who were lucky enough to work with George were forever shaped by his tenacious drive to understand the mechanisms underlying diverse biological phenomena. George often quoted Aharon Katchalsky, his postdoc advisor at Weizmann: *"It is easier to make a theory of everything, than a theory of something."* Modeling a biological system is indeed a daunting task and the "theories of something" that we have so proudly produced were often summarily invalidated by subsequent experiments. George, however, always insisted that despite the difficulties, we must focus on modeling specific systems because this modeling approach was more useful for advancing our knowledge of biology. While validation by subsequent experiments was certainly rewarding, contradiction motivated us to revise and improve our models. With each iteration, we learned something new and got closer to the truth. While we worked with George, we became disciples of his approach to science. Now, as independent investigators, we try to continue his legacy by exposing our trainees to his style of research. In what follows we recount what we believe are essential ingredients of George's success as a scientist.

**Lesson 1: Start from a mechanical picture**
Given the overwhelming complexity of biological systems involving a multitude of molecular components, where can theoreticians start in their modeling efforts? Some start with statistical analysis, looking for correlations in big data. Some build large-scale models with as much detail as possible in the system. Some focus on the molecular structures, and some on the entangled biochemical networks. George, in most of his best known works, focused on the mechanical aspects of a system, i.e., movements and shapes of objects directly observable in the experiments and forces acting on these objects.

What is the advantage of thinking about mechanics as a starting point in modeling biological systems? George pointed out that biological processes are tangible phenomena and as such are associated with mechanical actions. While the biochemical pathways behind a phenomenon can be very complex and hard to disentangle, the mechanical picture can be intuitively understood based on fundamental physical principles. Addressing mechanical questions in the system can hence provide a central framework to which additional biological details can be gradually added later on. Such a framework often brings about critical insights before even performing any mathematical analysis or numeric simulations. George always said that *"The art of science is to work on something doable while pushing the field forward"*; and mechanics had served as his chosen entry point to many, if not most, topics he worked on.

To illustrate how George started from a mechanical picture and how the resulting mechanical model served as a framework to incorporate complex molecular pathways, we recall his efforts on modeling endocytosis as an example. George began investigating this complex process by focusing on the changes of membrane shapes. As endocytosis progresses, the plasma membrane is invaginated into a tubule and eventually the endocytic vesicle buds off from the end through membrane fission. It was long presumed that the vesicle was pinched off by a shrinking collar of dynamin GTPases surrounding the membrane neck (a.k.a., the "pinchase" action). However, David Drubin and coworkers showed that dynamin GTPase is not essential for the final membrane fission in budding yeast endocytosis (Gammie *et al.*, 1995; Nothwehr *et al.*, 1995). If not the GTPase activity, what can drive this last step?

Having studied membrane mechanics starting in the 1980s (Jacobson *et al.*, 1986; Oster *et al.*, 1989; Kim *et al.*, 1998), George recognized that active force generation by proteins at the neck may not be necessary for membrane fission. He proposed that membrane fission could occur spontaneously as a result of mechanical instability. To explain the idea, he drew an analogy between the pinch-off of an endocytic vesicle and that of a soap bubble blown through a narrow ring. At the neck of a bubble the soap film is highly curved, and therefore, a tremendous amount of bending energy is highly localized. When the bubble neck is narrow, it may be mechanically unstable if the contribution of the negative curvature in the longitudinal direction overpowers the positive curvature in the radial direction. In this case, it is more energetically favorable for the neck to shrink (Fig.1A), which eventually leads to the bubble pinching off from the ring. The same mechanical instability could occur in the narrow neck of a budding membrane vesicle. This idea turned out to be correct and was later formally proven (Stephens *et al.*, 2017). Notably, mechanical instability does not have to drive the fission all the way -- when the vesicle neck shrinks to a few nanometers, thermal energy becomes sufficient to trigger spontaneous scission. In this conceptual framework, the protein machinery is only needed for invaginating the membrane; membrane mechanics and thermal fluctuations can then finish the job of membrane fission.

Based on this elegant concept, George asked a student, Jian Liu, to collaborate with the Drubin lab to build a theoretical model for the mechanism of endocytic membrane fission. After plugging into the model realistic parameters and the architecture of membrane invagination (Fig. 1B), Jian found that the membrane tubule – invaginated by the actin polymerization force – could not develop the sharply curved neck that was required for initiating the mechanical instability. This is because the strong tension along the tubule pushed in by actin filaments prevents curvature in the longitudinal direction, just as stretching a rubber sheet would make it more resistant to normal deformation. To solve this problem, George proposed an additional mechanism for curved neck formation. He argued that different proteins coating the tip and the stem of the membrane tubules could sort the lipid species and cause a phase separation at the membrane neck (differential colors in Fig.1B). The resulting line tension force could then make a narrow neck. Indeed, this additional factor made the model work and beautifully recaptured the membrane fission process (Liu *et al.*, 2006).

Building upon this framework grounded in membrane mechanics, Jian and George further constructed a more complex mechanochemical model, in which key endocytic players take their roles in a coherent mechanistic picture (Liu *et al.*, 2009). The mechanochemical model provided a comprehensive explanation of a precisely orchestrated spatiotemporal assembly and disassembly of all the essential proteins in endocytosis, which accompanies the development of membrane invagination and fission. Importantly, the model suggested that the robust sequence of endocytic events was orchestrated by an intricate coupling between membrane shape, mechanical force, and biochemical reactions. Of particular note, curvature-sensitive proteins, such as clathrin and BAR-domain proteins, are recruited by membrane curvature to distinct locations along the membrane invagination, and at the same time, deform the membrane locally. As such, the central mechanical framework coherently incorporated numerous components of a complex system, and pieced together a large amount of puzzling experimental observations. This modeling framework provides a useful roadmap for subsequent studies of endocytosis and related problems (Rangamani *et al.*, 2014; Hassinger *et al.*, 2017).

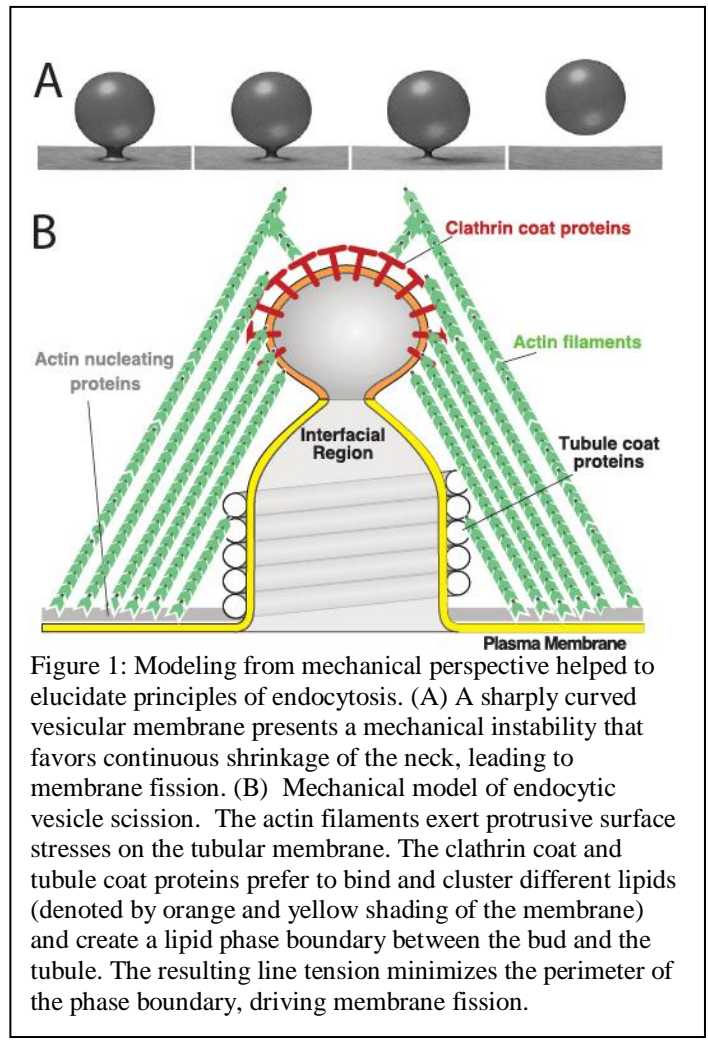

Figure 1: Modeling from mechanical perspective helped to elucidate principles of endocytosis. (A) A sharply curved vesicular membrane presents a mechanical instability that favors continuous shrinkage of the neck, leading to membrane fission. (B) Mechanical model of endocytic vesicle scission. The actin filaments exert protrusive surface stresses on the tubular membrane. The clathrin coat and tubule coat proteins prefer to bind and cluster different lipids (denoted by orange and yellow shading of the membrane) and create a lipid phase boundary between the bud and the tubule. The resulting line tension minimizes the perimeter of the phase boundary, driving membrane fission.

This is just one example of how **approaching a problem from the mechanical angle** is a hallmark of Oster-style biophysics. Addressing mechanical questions in the system provides an excellent starting point for further understanding. Of course, when thinking about mechanics at the cellular and molecular scale, one has to be careful to account for microscopic effects such as thermal fluctuations. Once the mechanical framework is constructed, it can later be supplemented by the molecular details resulting in a comprehensive mechanochemical model of the phenomena. Such models utilizing mechanics at their core have, and will continue to, provide answers to challenging puzzles in biophysics.

**Lesson 2: Do not be afraid to be wrong, but be constrained by data**
Finding the right mechanistic or mechanical picture for a wide range of biological systems requires one to think outside the box. This is especially important when experimental data are scarce and fragmented and do not lead to a clear mechanistic picture. This is exactly when theoretical models bring the greatest values to experimental research -- by providing testable predictions that are otherwise unobtainable by intuition. In these situations, the key was to come up with a simple hypothesis consistent with all the constraints set by the experimental observations and by fundamental laws of physics. While the model hypotheses may only take up a few lines in the published articles, generating these hypotheses was actually the rate-limiting step of our research. This was where George's ingenuity shined and where he trained us to think outside the box.

To cultivate a brainstorming atmosphere and stimulate the free flow of ideas, George got us out of the lab, and better yet, off campus. Every morning, five days a week, George conducted his group meetings in a nearby coffee shop. There were three rules to these meetings: bring a pen, a scratchpad, and an open mind.  As we sat at a table, perhaps immediately next to a group of fellows from the nearby Berkeley Divinity School discussing theology, we immersed ourselves in the idea-generating mode. The most critical part of these discussions lied in drawing from seemingly unrelated systems and use these parallels to formulate the hypotheses relevant to our systems. Once a hypothesis was generated, it was time for us to put our "skeptic" hat on and try to argue about the validity of the hypothesis based on what we knew about the underlying physics and biology. As with any brainstorming session, many of the ideas turned out to be wrong, contradicting the established biological facts or even the basic laws of physics. But George was never afraid to propose bold and crazy ideas, and inspired us to do the same. Over several rounds of failures, the one ingenious idea that could withstand all the tests emerged. As Alex Mogilner, one of George's postdocs and collaborators would later quip, *"George would come up with an idea that you would think was completely crazy. You would go off and work on the project using a different hypothesis, and then, after a number of months, you realized that George's crazy idea could actually work and was even likely correct."*

To illustrate the process, some of us recall the brainstorming on how biofilms containing millions of cells of *Myxococcus xanthus*, a soil bacterium, self-organize into regular-spaced waves. These waves, termed ripples, travel in space as bands of high-density crests surrounded by lower density troughs (Reichenbach, 1965; Shimkets and Kaiser, 1982; Welch and Kaiser, 2001). Despite their discovery in the mid-60s by Reichenbach and colleagues (Reichenbach, 1965), the mechanism by which the waves form was not clear when George assigned this project to Oleg Igoshin, a graduate student who had just joined the lab in the fall of 2000.

A key question in constructing a model for *M. xanthus* waves was how an initially homogeneous distribution of cells could self-organize into traveling bands. Coffee-shop discussions generated a multiplicity of ideas that were quickly ruled out. The first analogy was drawn from diffusion-reaction waves in chemical systems such as arise from the Belousov-Zabotinsky (BZ) reaction (Winfree, 1984). Could chemical substances secreted by the cells form such a system? Or could the bacterial cells switch between states with different diffusion coefficients (presumably reflecting different motility states) like the chemical species in the BZ system and hence result in the traveling wave patterns? These ideas were quickly ruled out as it was hard to relate to the experimentally observed behavior of cells. Furthermore, as George noted, the diffusion-reaction wave crests like those in the BZ system would annihilate each other upon collisions, rather than passing through one another as the *M. xanthus* ripples do. A more fruitful analogy came from models of the fruit fly circadian clock (Winfree, 1970) and the clock-and-wavefront models of segmentation in vertebrate development (Jiang *et al.*, 2000; Baker *et al.*, 2006). Given that each *M. xanthus* cell is known to periodically switch its polarity and reverse its direction, George hypothesized that these reversals could be controlled by an underlying oscillator or clock.

Synchronization of these clocks in space and/or time could lead to ripples, as long as there was a mechanism that could synchronize the individual cellular clocks. Intriguingly, the reversal frequency of *M. xanthus* cells was suggested to be modulated via cell-cell contact signaling (Jelsbak and Sogaard-Andersen, 1999). In one of the coffee sessions, Oleg (who grew up in snowy Siberia) came up with another useful analogy – the wave-crest acts as a snow-plow. If cells in the crests signal counter-moving cells to reverse and join them, the crests would accumulate cells over time. George liked the analogy and suggested an additional model ingredient – a refractory period – once reversed cells cannot be ready to reverse for some time and, therefore, stay in the crests. When put in mathematical form these hypotheses explained all the known wave properties (Igoshin *et al.*, 2001). The prediction of space-time synchronization in the ripples was later confirmed by experiments (Welch and Kaiser, 2001; Zhang *et al.*, 2012).

Another secret to George's success as a biophysical theorist was his incessant pursuit to further challenge his models and theories with new observations. Even when we had a model that solved the original puzzle and explained the existing data, George was always on the lookout for additional data that could further challenge or constrain the model. He liked to quote from Katchalsky, *"Theory tells us what cannot happen, and it can tell us what could happen. But only experiments tell us what does happen."* He would never hesitate to call up or email an experimentalist for additional data to test the predictions or assumptions of his models. This practice of tightly linking the theories and models with experimental data and reiterating the modeling became a research philosophy of a majority of the group's alumni.

For example, when working on the first model of *M. xanthus* gliding motility, Charles Wolgemuth, then a postdoc with George, was motivated by a paper by Egbert Hoiczyk and collaborators proposing that motility of cyanobacteria chains could be driven by slime secretion (Hoiczyk and Baumeister, 1998). Given that bacterial cells live at ultralow Reynolds numbers where force generation by jet-propulsion is not possible, they had to find an alternative biophysical explanation for how force could be generated by slime secretion. The mechanism proposed by Charles and George was inspired by a machine previously built by Katchalsky and collaborators (Katchalsky and Lifson, 1954). The machine generated mechanical work through expansion of polyelectrolyte strips driven by changes in osmolarity or pH. Similarly, in the "slime-gun" model that Charles developed, polyelectrolyte gel (slime) produced in the high-osmolarity environment inside the cell is secreted into a small nozzle and expands as it moves to the low-osmolarity external environment. As the gel expands, it pushes back on the *M. xanthus* cell, thereby generating thrust. Given the estimated force from a single nozzle, the model predicted the number and geometry of slime-secreting pores required in each cell to generate a sufficient propulsive force to account for the observed cell velocity. These hypothetical pores, though, had not been observed in *M. xanthus.* Within a week, George reached out to Egbert, who was a postdoc at that time at The Rockefeller University working on an unrelated project, and convinced him to perform electron microscopic characterization of the slime-secreting pores in *M. xanthus*. The resulting collaboration not only proposed the first biophysically realistic model for bacterial gliding motility (Wolgemuth *et al.*, 2002), but also changed the career trajectory of Egbert, who began working on *M. xanthus* once he started his own lab. While the slime-gun model was later disproved for *M. xanthus* (Sliusarenko *et al.*, 2007; Nan *et al.*, 2011; Faure *et al.*, 2016)*,* it still may hold for the cyanobacterial cell chains.

To summarize, the ability to generate, filter and refine ideas that can solve biological puzzles is the cornerstone of Oster-style biophysics. He cultivated these abilities in all of his trainees, and we all forever remain grateful for these lessons.

**Lesson 3: Be the first one to correct your models**

As much as he embraced crazy ideas, George remained a critical reviewer of his own ideas even long after they were published. George particularly emphasized a neutral mind towards one's own work and a mental readiness to revise it or shoot it down as new and contradicting evidence emerges. He frequently stressed the importance of striking the right balance between persistence and open-mindedness. On one hand, we should have the courage to defend the ideas we believe to be right. On the other hand, we should not, as he joked about it, "marry a model till death do us part"*,* but should rather frequently reexamine it in light of new evidence as a critical reviewer would do. He emphasized that making mistakes is not shameful, but refusing to accept them is. Over and over again he liked to say, *"If your model is going to be proven wrong, you should be the first one who does it."*

As a good example, George's persistent efforts to address new experimental evidence led to three different models for the bacterial flagellar motor (BFM). The BFM drives flagella-mediated swimming motility in many bacteria such as *E. coli* (Berg, 2003). The BFM is a large protein complex embedded in the membrane, which garners energy from the transmembrane ion ($H^+$ or $Na^+$) gradient in order to rotate the flagellum. An intriguing question is how the transmembrane ion flux powers the relative rotation between the rotary and stationary motor parts. George's attempt to answer this question dated back to 1997. Together with his postdoc, Tim Elston, George developed a BFM model reminiscent of a turbine engine (Elston and Oster, 1997). In this model, ions jump on and off negatively charged stator sites located at the interface between the rotor and stator from either the periplasmic space or the cytoplasm. A relative tilt between the charged sites on the stator versus those on the rotor couples the translocation of ions through the membrane to relative rotation between the rotor and stators, similar to how tilted blades convert face-on wind into rotation of a turbine engine. The energy-driven net flux of ions from the cytoplasmic side to the periplasmic side causes rotation of the rotor in the coupled direction. The model successfully explained most of the experimental observations at that time.

However, subsequent structural studies showed a physical separation of the ion translocation path from the rotor-stator interface. This structural information indicated that the rotor is not directly involved in the ion translocation process, calling into question the main premise of the conceptually appealing protein turbine model. In addition, George was also bothered by the inability of the model to explain a biphasic transition in the slope of the experimentally measured torque-speed curve (Fig. 2A). George insisted that either there was an artifact in the experimental results, or something was missing from the model. To resolve these discrepancies,

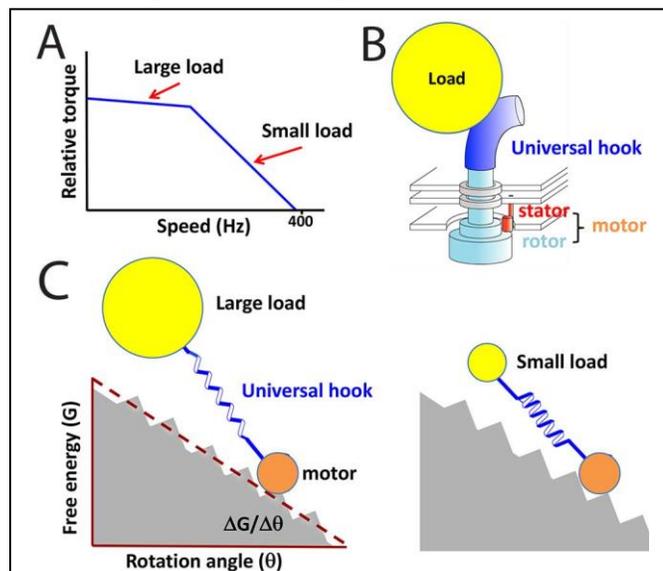

Figure 2: Simple theory explains torque-speed relation in BFM. (A) Experimentally observed torque-speed relation displays sharp increase in the slope at a threshold speed. (B) In the experiments, the load was applied through the universal hook of the BFM, which acted as an elastic spring. (C) Illustration of the major conclusion of the model by Xing et al. (2006). The motor cycles can be illustrated as a particle moving down a rugged energy landscape. The elastic universal hook buffers the motion of large loads and results in temporally averaged smooth free energy surface. For such sluggish loads (or equivalently for low speeds) the maximum torque is determined by the thermodynamic driving force, $\Delta G/\Delta\theta$. For small loads or at high speed, the buffering is weak and the resulting rugged landscape causes some of the mechanical energy from the motor to be dissipated directly as heat. This dissipation leads to decrease in torque above ~300 Hz.

George asked his postdoc, Jianhua Xing, to collaborate with Richard Berry from Oxford and his student, Fan Bai, to construct a new BFM model. Based on the newly available structural and biochemistry evidence, the team proposed an alternative mechanism to explain how ion translocation is coupled to motor rotation (Xing *et al.*, 2006). Basically, binding and unbinding of ions to their binding sites on the stator induces a conformational change in the stator that drives rotor rotation. In other words, the stators in this model work similarly to kinesin or myosin stepping along tracks, except that the track here is not a linear microtubule or actin filament, but rather the circular rotor. The team also provided a simple physical explanation for the torque-speed relationship (Fig. 2A). The key was to notice that a universal flagellar hook (Block *et al.*, 1989) serves as a soft spring connecting a motor to its load (Fig. 2B). The soft hook would effectively integrate out the temporal fluctuations in torque under high-load/low-speed conditions (Fig. 2C). Essentially, the high-load/low-speed scenario represents a quasi-equilibrium process analogous to the thermodynamics textbook example of infinitesimally slow expansion of ideal gas in a piston-cylinder device. The load corresponds to the piston, and the motor stepping corresponds to the motion of gas molecules. The quasi-equilibrium process allows maximal work to be generated, which is determined by the free energy drop in the system, $\Delta G$. The low-load/high-speed scenario, in contrast, cannot convert all the free energy difference to mechanical work and hence generate lower torque. This theory is generic and applicable to any motor systems with an elastic linker structure that connects to cargo and further demonstrated the power of a mechanical perspective. It led to insights that would not have been obvious if one had followed the established biochemical models and treated the motor cycle as transitions among discrete chemical states.

Despite these advances, but not surprisingly, George continued to refine the BFM model as new data became available. In 2015-16, George and coworkers published a third major version of the model, which further incorporated the latest structural details and showed how ion binding and unbinding affect protein interactions and drive motor rotation (Mandadapu *et al.*, 2015; Nirody *et al.*, 2016). This was George's last work on protein motors. If he were still with us, he would likely continue to think about these systems and critique his own models as new data emerges.

As willing as George was to revise his theories in light of new data, he was also ready to challenge the conclusions of experimental papers based on insights from his models. He liked to paraphrase Francis Crick, *"The model should not fit all the data because not all the data is correct."* A good example was shown in George's efforts to model another rotary motor, the F1Fo ATP synthase. The cytosolic F1 and transmembrane Fo parts are two rotary protein motors mechanically coupled to each other through a central rotary shaft, working to interconvert between ATP production and transmembrane proton or sodium transport. In their earliest models for Fo (Elston *et al.*, 1998; Dimroth *et al.*, 1999), George and coworkers came up with a model that successfully explained most of the observations at the time. Similar to the flagellar motor, new structural information questioned the original model assumptions. In light of the new data, he asked Jianhua Xing to resume the collaboration with Peter Dimroth, an experimentalist at ETH Zurich, to revise the earlier model. Through what George would call "mutual education and iterative pruning" (Drubin *et al.*, 2010), the team developed a new Fo model that incorporated the new structural data, and made predictions that were confirmed by the Dimroth's group (Xing *et al.*, 2004). One of Jianhua's predictions, however, contradicted observations in Dimroth's lab. George and Jianhua insisted that the prediction was a necessary model outcome derived from Peter's previous experimental paper. Their persistency led Peter to discover a mistaken unit conversion and consequent faulty conclusion in the original experimental study.

Thus, painstaking revisions of the earlier models was another hallmark of George's career. The revisions were motivated by experiments and pointed out new directions for experiments. The iterative cycles of modeling and experiment allowed George and his collaborators to gradually perfect their explanations to the puzzles that biology presents.

**Lesson 4: Convey your model to biologists**
Convincing experimentalists to test model predictions is a hard task that requires excellent communication skills. For a biophysical model or theory to have real impact on the field, its presentation should be accessible to the broader biological community. George was not a big fan of theory papers written by theorists for theorists. He always aimed to publish papers in journals that experimentalists read regularly and in a form that was easily accessible to them. This is the style of communication that he passed on to his trainees.

George stressed that while we need math to confirm the idea and to define its validity, we don't necessarily need math to explain the idea. He made us describe all the model ingredients with the fewest equations possible in the main text and to leave detailed derivations and formal descriptions to the methods and supplemental sections. The same lesson also applies to delivering conference talks, especially for an audience of mostly experimental biologists: focus on the ideas that are relevant to the biology, not theoretical formalism.

For the same reasons, George could never emphasize enough how important it was to create intuitive graphical representations of our models and results. He was proficient in graphical illustration software and ingenious in capturing complex model ingredients and results with simple drawings. To emphasize the importance of graphical illustrations of the results, George shared a lesson he learned from his earlier study of seashell patterns. Collaborating with Bard Ermentrout,

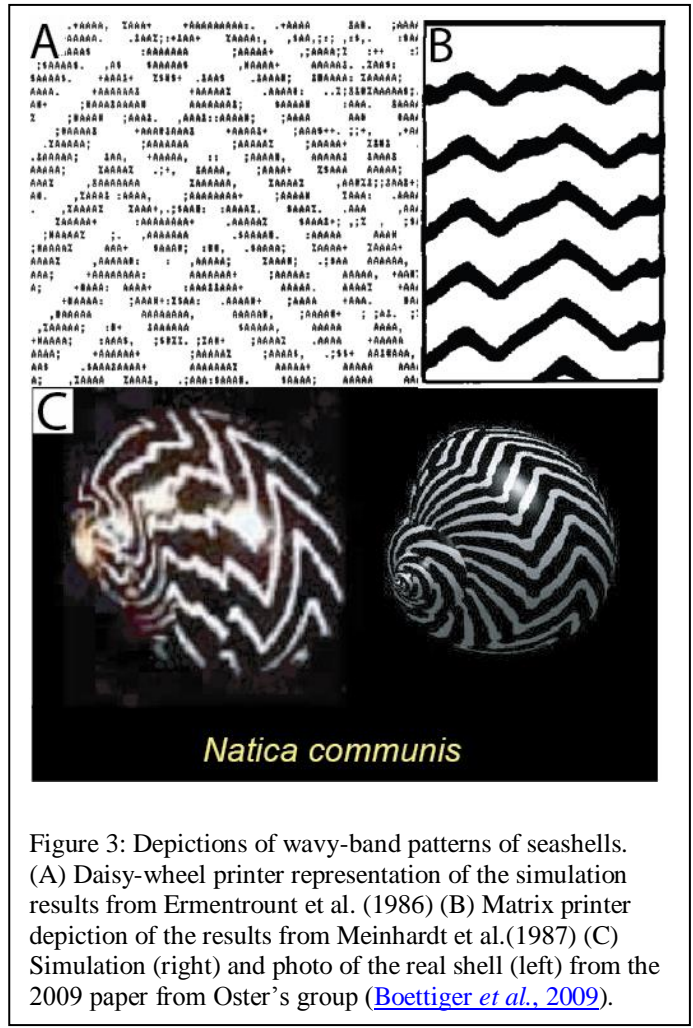

Figure 3: Depictions of wavy-band patterns of seashells. (A) Daisy-wheel printer representation of the simulation results from Ermentrount et al. (1986) (B) Matrix printer depiction of the results from Meinhardt et al.(1987) (C) Simulation (right) and photo of the real shell (left) from the 2009 paper from Oster's group (Boettiger *et al.*, 2009).

George published the first biologically realistic model of seashell patterns (Ermentrout *et al.*, 1986), about one year prior to the publication of a competing model by Meinhardt and collaborators (Meinhardt and Klingler, 1987). Arguably, Meinhardt's model was less realistic as it associated the pattern formation with diffusible morphogens that were never identified. The Meinhardt paper, however, became much more widely accepted and cited. George attributed part of the popular success of Meinhardt model to their use of the then state-of-the-art matrix printer (Fig. 3B), which made the illustrations superior to the ones made by George and Bard, who only had access to a daisy wheel printer (Fig. 3A). Consequently, when George revisited the problem with a new theory in 2009, the illustrations of the results were worthy of display at natural science museums (Fig.3C) (Boettiger *et al.*, 2009).

**Lesson 5: People matter more than projects**

George was not only a great scientist, but a great soul. He always emphasized that *"people matter more than projects"*. He often said that his secret of success is to *"work with people who are smarter than you, and who know things you don't."* This mindset was the foundation upon which he was able to synergize his own ingenuity with other great minds to generate groundbreaking theories about biology.

George was one of the most open-minded scholars in the world. He loved the experimental biologists coming to him with their puzzles and was never too shy to call or email a colleague to pick their brain on a topic of their expertise. Many such discussions blossomed into successful collaborations and life-long friendships. The collaboration with the David Drubin, a renowned cell biologist, on endocytosis, was a great example, which lasted more than a decade until George passed away. This experience was so successful that they published a commentary together to share their thoughts about collaborations between experimentalists and theorists (Drubin *et al.*, 2010). On the other hand, George also often teamed up with applied mathematicians who helped bring rigor and analytical insights to our models. For example, John Neu, at the time a professor of mathematics at Berkeley, became a great friend and collaborator of George, and this collaboration lasted for more than a decade. John was a frequent visitor to George's morning coffee hours, where they entered into lively discussions of their projects. They benefited each other with their complimentary expertise, John with superb mathematical skills and George with broad knowledge and acute intuition about how different parts connect and function in biological mechanisms. Over the years their collaboration led to seven publications on protein-membrane interactions, myxobacterial pattern formation and bacterial gliding motility (Kim *et al.*, 1998, 2000; Grabe *et al.*, 2003; Igoshin *et al.*, 2004; Sliusarenko *et al.*, 2006; Chen *et al.*, 2009; Nan *et al.*, 2011). In 2009, the Oster-Inspired Research Conference was held in Berkeley to celebrate George's life contributions to science. Many of his former and ongoing collaborators attended the conference, and they represent an amazingly broad range of scientists.

George paid respect and appreciation to the intellectual power of not only his peer scientists, but also his trainees. This made him a truly great mentor and adviser. He encouraged us all to partake equally in discussions. No matter what cultural background we came from, everyone was able to quickly engage into the heated and free scientific debates in the lab. "As a Chinese girl,"

Jing Chen reflected on her initial interactions with George as a graduate student, "I was brought up and taught to be polite and obedient to authorities. During my first year in the lab, I felt guilty and apologized to him whenever I had a heated, 'impolite' argument with him. He laughed at my apologies and told me to keep arguing with him about science. He really taught me how to enjoy the pure fun of science." Through open and frank critique of each other's and George's ideas, we grew into mature scientists with sharp minds and confidence. We all thank George for creating such an unassuming atmosphere where stubbornness and foolishness can give birth to wonders.

George was a true practitioner of equity, diversity, and inclusion, by always showing support for his trainees and placing implicit trust in their ability to maintain a balance between work and life. Around 2010, two of his postdocs were young mothers of two children. "When I joined his lab, I had a 3-year old and a 3-week old at home," said Padmini Rangamani, his former postdoc, "I was quite unsure of how my personal situation would be perceived in terms of my commitment to science. George was quick to put me at ease during my interview, sharing that he had a child and that he got that on some days, the kids would simply need more of my time. More importantly, during my 4-year training with him, he never once made me feel like I should be working more or differently. His implicit trust in my ability to manage the demands on my time and that I'd use the flexibility that a postdoc offers to the best of my abilities is one of the main reasons I was able to do as well as I did and subsequently applied for faculty positions."

Last but not least, George's openness and generosity laid another cornerstone for his success in training of young scientists. George was always happy to let his trainees pursue their own interests and passion. Although he contributed substantially to the maturation of these research topics in many cases, George never felt territorial about his ideas. He was very happy to let his trainees carry their projects off to their independent positions. At the same time, the creative atmosphere and the broad training George provided allowed many of his lab alumni to quickly move into and succeed in different fields. Those of us who now lead independent groups are conducting research in a wide range of topics, from molecular biophysics and cellular mechanisms to population dynamics and neuroscience. Quoting from Alex Mogilner's speech at George's memorial service, "The work carried out by each of us look like footnotes to George's chapter in the book of science. Because this chapter is so wonderful, even the footnotes look pretty good."

**Concluding remarks**

As a scientist, as a mentor and as a friend, George served as a great role model for people around him. No matter which cultural or educational background we came from, everyone in George's lab was able to quickly engage into proposing, discussing, critiquing, and defending topics that caught their eyes. In retrospect, many would agree that their time with George is one of the happiest and most important periods in their career development, and his influence is life-long. George's unceasing curiosity, coupled with his deep caring for others, culminated in some ten thousand discussions at the coffee shops with his trainees, collaborators – and most importantly, friends. It was ultimately George's humanity that inspired and sustained a vibrant group of scientists that have made it their life's goal to unravel the mysteries of nature.

The approach to biophysics and computational biology that George cultivated in his group led to many paradigm-shifting models. To George and his trainees, modeling is an intellectual journey that starts with data, winds along with imagination bound by physical and chemical laws, and finally returns to reality – a journey of constantly seeking the beauty of truth by constantly correcting ourselves. The resulting models were significant, often not because they give the ultimate answers, but because they brought critical insights and unprecedented perspectives to puzzling and fragmented experimental observations. In this way, these models illuminated the path of subsequent scientific endeavors.

As computational biology matures as a field the amount of available data multiplies and computational power grows, it has become trendy to outsource idea generation and verification to a computer chip through large-scale modeling and statistical analysis. While the big-data approaches undoubtedly have their niche, we believe that Oster-style biophysics will continue to be fruitful in solving the puzzles nature throws at us. Rather than hoping for computation to tell us the answer, George leveraged his wide breadth of knowledge of the biological literature to formulate mechanistic -- or better-yet mechanical -- hypotheses that suggest plausible solutions. Through the open atmosphere and heated discussions, hypotheses were refined or discarded. Then it was time to translate these ideas into equations and check their consistency with available data and the laws of physics. When a critical piece of data was missing, George strived to establish collaborations to obtain it. In the end, the papers were always honest about assumptions and limitations, while striving for a graphical, easy-to-grasp explanations of the findings. The papers also proposed cornerstone experiments that could confirm or kill the theory. In the latter case, George was always happy to go back and work out an alternative solution. With each iteration of modeling and collaborative experimental testing, we got ever closer to the ultimate solutions of the biological puzzles.

## Acknowledgements

We thank Dr. David Drubin for his encouragement and help in streamlining the review in the context of the MBoC special issue: "Mechanical forces on and within cells". We also thank Drs. John Neu, Alex Mogilner and Jung-Chi Liao for their useful suggestions and comments at various developmental stages of the manuscript. We are grateful to Alistair Boettiger and Bard Ermentrout for providing materials for Figure 3. The work was supported by National Science Foundation awards MCB-1616755 (to O.A.I.) and DMS-1462049 (to J.X.) and by start-up fund from John Hopkins University (to J.L.).